\date{}
\documentstyle[prl,aps,preprint,epsf]{revtex}

\begin{document}
\draft 
\title{Random Analytic Chaotic Eigenstates} 
\author{\large P. Leb{\oe}uf}
\address{Laboratoire de Physique Th\'eorique et Mod\`eles Statistiques 
  \footnote{Unit\'e de recherche de l'Universit\'e de Paris XI associ\'ee au 
  CNRS}, B\^at. 100, \\ 91405 Orsay Cedex, France}

\maketitle
\begin{abstract}
The statistical properties of random analytic functions $\psi (z)$ are
investigated as a phase-space model for eigenfunctions of fully chaotic
systems. We generalize to the plane and to the hyperbolic plane a theorem
concerning the equidistribution of the zeros of $\psi (z)$ previously
demonstrated for a spherical phase space ($SU(2)$ polynomials)). For systems
with time reversal symmetry, the number of real roots is computed for the
three geometries. In the semiclassical regime, the local correlation
functions are shown to be universal, independent of the system considered or
the geometry of phase space. In particular, the autocorrelation function of
$\psi$ is given by a Gaussian function. The connections between this model
and the Gaussian random function hypothesis as well as the random matrix
theory are discussed.
\end{abstract}


\newpage

\section{Introduction}

There are two basic descriptions of the eigenfunctions of fully chaotic
systems. On the one hand, there is the random matrix theory of Wigner and
Dyson as a model of the quantum behavior of chaotic motion
\cite{mehta,bohigas89}. In this model the Hilbert space is isotropic, i.e. an
arbitrary eigenstate $\varphi$ of the random matrix can point out in any
direction of Hilbert space with equal probability. The only constraint comes
from the normalization. Denoting by ${\vec a} = (a_0, a_1, \cdots , a_N)$
the components of $\varphi$ in a given (and arbitrary) basis, the joint
probability distribution for the amplitudes is \cite{porter}
\begin{equation} \label{1} 
 {\cal D}_{RMT} ({\vec a}) = \frac{1}{| S_{2(N+1)} |}
\delta \left[\sum^N_{m=0} |a_m |^2 - 1 \right] \ ,
\end{equation}
where $|S_n| = 2 \pi^{n/2}/\Gamma (n/2)$ is the surface of a
$(n-1)$-dimensional sphere of unit radius. We assume for the moment that
${\vec a}$ is a complex vector, this being the usual situation when the
physical system has no time-reversal symmetry. By construction this theory
concentrates on statistical aspects of the eigenstates, and the
"ergodicity" in Hilbert space expressed by Eq.(\ref{1}) is a direct
consequence of the invariance of the ensemble of matrices under unitary
transformations.

On the other hand, a semiclassical theory has also been developed 
\cite{voros76,berry77,bogo88,berry89}. Unlike the random matrix theory,
its aim is to provide a system-dependent description of the eigenstates
based on the solutions of the classical dynamics. Adopting a phase-space
representation of quantum mechanics, in the semiclassical limit $\hbar
\rightarrow 0$ the Wigner transform ${\cal W}_\psi ({\vec x},{\vec p},E)$ of an
eigenstate $\psi$ of an ergodic system at energy $E$ takes to leading order
the simple form 
\cite{berry89}
\begin{equation} \label{2}
{\cal W}_\psi ({\vec x},{\vec p},E) \approx {\cal N} \delta\left[ E-H({\vec x}
,{\vec p})\right] \ ,
\end{equation}
where ${\cal N}$ is a normalization constant, $({\vec x},{\vec p})$
parametrizes the phase space and $H({\vec x},{\vec p})$ is the classical
Hamiltonian. Because in the semiclassical limit the Wigner function presents
wild oscillations, Eq.(\ref{2}) has to be interpreted as a smoothed version
over a phase space region of the order of one Planck volume.

Besides the energy-shell ergodicity, Eq.(\ref{2}) reveals a strong
localization effect in phase space of which there is no trace in
Eq.(\ref{1}). Because of energy conservation ${\cal W}_\psi$ is, in the
semiclassical limit, exponentially concentrated on the energy shell. This
means that when expanding an eigenstate of $H$ in an arbitrary basis
$\phi_m$ of Hilbert space it will have components different from zero only
on those states whose Wigner transform intersects -- or is included into --
the energy shell $\delta\left[ E-H({\vec x},{\vec p})\right]$, while the
amplitudes will vanish exponentially fast otherwise. Therefore, in general
for a given chaotic Hamiltonian system the expected distribution of the
eigenstates's amplitudes is quite different from the isotropic rule
(\ref{1}), because random matrix theory ignores the phase-space restrictions
imposed by energy conservation. Eq.(\ref{1}) has therefore to be interpreted
as a valid statement only for the amplitudes $a_m$ whose associated vector
$\phi_m$ intersects the energy shell.

Having this in mind, our purpose is to further develop the connections
between the two above mentioned descriptions. In particular, assuming a
two-dimensional phase space we study the ergodicity of the eigenstates as
implied by Eq.(\ref{1}), as well as some correlation functions computed
inside the energy shell. For that purpose, we use the Bargmann (or coherent
state) representation which associates to each quantum state an analytic
function $\psi (z)$ in the complex plane. This complex plane is interpreted
as a parametrization of the energy shell manifold, and the three simplest
one-dimensional complex manifolds are considered: the sphere (curvature
$+1$), the plane (zero-curvature) and the hyperbolic plane or pseudosphere
(curvature $-1$). After introducing the model in section II, we show in
section III that the associated Bargmann functions given by Eq.(\ref{3})
below have, when the coefficients $a_m$ are distributed according to
Eq.(\ref{1}), a uniform distribution of zeros over the sphere, the plane and
the hyperbolic plane, respectively. The uniformity of the distribution was
shown in \cite{bbl1,bbl2} for the spherical geometry, and the
results of this paper constitute a generalization to the plane and to the
hyperbolic plane. Some related theorems extending some of the results to
higher dimensional phase spaces were recently obtained independently
in Ref.\cite{sz}.

We also consider the case when the $W_1$ and $SU(1,1)$ series in
Eq.(\ref{3}) are truncated to a finite number of terms (thus introducing
$W_1$ and $SU(1,1)$ polynomials). It is shown that for those polynomials the
distribution of roots is now uniform over a finite region (a disk) of the
complex plane and the hyperbolic plane, respectively.

The equidistribution of roots holds only if the vector $\vec{a}$ is complex.
When it is real (which corresponds to time-reversal symmetric systems), a
concentration of roots occurs over the real axis in the complex plane. 
We show that for the three random functions considered the number of real
roots located inside a disk centered at the origin containing $N$ zeros  
is proportional to $\sqrt{N}$.

Section IV is devoted to the study of the correlation properties of the
random analytic functions between different points of the energy shell. We
first argue that in the semiclassical limit all the correlation functions
are locally universal, independent of the geometry of phase space. We then
compute as a particular example the autocorrelation of $\psi (z)$.

\section{Random analytic functions}

In order to study the statistical properties of eigenstates of chaotic
systems, we first define an energy-shell (phase space) associated to the
quantum mechanical system. We will not consider a specific dynamical system
but make general statements which are expected to be valid for any system
having at the classical level a strongly chaotic behavior.

Concerning the energy shell, the minimum dimensionality required in order to
have chaotic behavior at the classical level is two. This two-dimensional
manifold ${\cal M}$ could be for example the Poincar\'e surface of section
of the energy shell of a two-freedoms Hamiltonian system. But we can
alternatively choose a two-dimensional manifold and directly define on it a
fully chaotic area-preserving map, thereby avoiding the problem of the
reduction to the Poincar\'e section. In both cases one can define a unitary
operator which quantizes the classical flow. In the following we study
properties of the eigenstates of this unitary operator. Because in principle
it corresponds to a section of it, we will still call the manifold ${\cal M}$
the ``energy shell''.

Independently of the definition of the dynamics, the three simplest
possibilities for ${\cal M}$ are, in two dimensions, the sphere, the plane
and the hyperbolic plane or pseudosphere \cite{bv86} (the cylinder and the
torus being just periodicized versions of the plane), having curvature $+1$,
$0$ and $-1$, respectively. The choice of the manifold will depend on the
dynamical symmetry group underlying the system Hamiltonian under study. The
three spaces considered are associated with the $SU(2)$, $W_1$ and $SU(1,1)$
symmetry, respectively. For example, for the motion of a spin in a variable
magnetic field the appropriate phase-space manifold is the sphere (see
Ref.\cite{perel} for illustrations in different geometries). Area preserving
chaotic maps have been studied in the plane and its periodicized versions
(e.g. the cat, standard and kicked Harper maps) and the sphere (the kicked
top) but not, to our knowledge, in the pseudosphere.

The next step is to define a representation of the quantum states over the
manifold ${\cal M}$. This can be achieved for the three geometries under
consideration by means of a phase-space coherent state (Bargmann)
representation of quantum mechanics consisting of overcomplete and
non-orthogonal systems of Hilbert space vectors. An arbitrary quantum state
can be written in the coherent state basis as
\cite{barg,perel}
\begin{equation} \label{3} 
\psi (z) = \left\{ \begin{array}{ll} 
\sum\limits^N_{m=0} \sqrt{ C^m_N} \, a_m \, z^m \ , \;\;\;\;\;\;\; &
SU(2) \\ 
\sum\limits^\infty_{m=0} \frac{1}{\sqrt{m! \hbar^m}}
\, a_m \, z^m \ , & W_1 \\
\sum\limits^\infty_{m=0} \sqrt{C^m_{L+m-1}} \, a_m \, z^m \ ,
\;\;\;\; & SU(1,1) \end{array} \right.
\end{equation}
where $C^l_k = k!/l! (k-l)!$ are Newton's binomial coefficients. For $SU(2)$
we have $N=2 j$, where $j=1/2, 1, 3/2, \cdots$ is the usual spin label of
the $(2 j +1)$-dimensional representation. For $SU(1,1)$, $L=2 k$, with
$k=1, 3/2, 2, \cdots $ is the usual label of the discrete representation of
the group \cite{perel}. Unlike the case of $W_1$ and $SU(1,1)$, the
representation of the quantum state for $SU(2)$ is finite dimensional,
reflecting the fact that the group is compact. 

Eq. (\ref{3}) provides a realization of the Hilbert space in terms of
analytic functions $\psi(z)$. The complex variable appearing in
Eq.(\ref{3}) spans in each case the associated manifold ${\cal M}$. It is
defined as
\begin{equation} \label{proj} 
z = \left\{ \begin{array}{ll}
\cot (\theta/2) \, \exp\left( {\rm i} \varphi \right) \ , \;\;\;\;\;\;\;\; & SU(2) \\ 
(x - {\rm i} p )/\sqrt{2} \ , & W_1 \\
\tanh (\tau /2) \, \exp\left( {\rm i} \varphi \right)
\ , & SU(1,1) \ . \end{array} \right.
\end{equation}
Here $(\theta,\varphi)$, $(x,p)$ and $(\tau,\varphi)$ are the usual
variables labeling these manifolds (see Fig.1). The measure of these three
manifolds projected onto the $z$-plane is
\begin{equation} \label{mu}
\mu (z,{\bar z}) = \left\{ \begin{array}{ll}
\displaystyle \frac{1}{\pi} \frac{d^2 z}{(1+z {\bar z})^2} \;\;\;\;\;\; & SU(2) \\
\displaystyle   d^2 z/(\pi\hbar) & W_1 \\
\displaystyle \frac{1}{\pi} \frac{d^2 z}{(1-z {\bar z})^2} & SU(1,1) \ .
\end{array} \right.
\end{equation}
The notation $d^2 \omega$ where $\omega$ is a complex variable stands for $d
({\rm Re} \omega) d ({\rm Im} \omega)$.

Now comes the statistical ingredient entering in the definition of the
model. If $\psi$ is an eigenstate of the quantization of the map acting on
${\cal M}$ then, for a fully chaotic dynamics, we assume that the amplitudes
$\left\{ a_m \right\}$ in Eq.(\ref{3}) are distributed according to
Eq.(\ref{1}). Of course this is an hypothesis whose validity has to be
compared with the behavior of real dynamical systems. Including this
element, we obtain therefore a phase-space model for chaotic eigenstates
in terms of {\sl random analytic functions}. For $SU(2)$ polynomials this
model was introduced in Ref.\cite{bbl1} and further considered in
\cite{ek,hann,bbl2,ls,bleher}.

Being analytic, the average properties of the random functions $\psi (z)$
cannot be stationary over the complex plane (unless the function is
trivially a constant). For example, for $SU(2)$ the average square
modulus of $\psi (z)$ grows like
\begin{equation} \label{avm} 
\langle \psi(z) {\overline{\psi(z)}} \rangle = \sum_{m,n=0}^{N} \sqrt{C_m^N
C_n^N} \langle a_m {\bar a}_n \rangle z^m {\bar z}^n = (1+z {\bar z})^N \ .
\end{equation}
The brackets indicate average over a uniform distribution for the
coefficients, $\langle f ({\vec a}) \rangle = \int d^2 a_0 \ldots d^2 a_N
{\cal P}({\vec a}) f ({\vec a})$. For convenience we use for the computation of the
averages a different normalization from that of Eq.(\ref{1})
\begin{equation} \label{pa} 
 {\cal P} ({\vec a}) = \frac{2 (N+1)}{| S_{2(N+1)} |}
\delta \left[\sum^N_{m=0} |a_m |^2 - (N+1) \right] \ .
\end{equation}
We therefore have 
\begin{equation} \label{5}  
\langle a_m {\bar a}_n \rangle = \delta_{mn} \ ,
\end{equation}
from which we obtained Eq.(\ref{avm}). This change of normalization has no
incidence in the results reported in this paper since the quantities
considered (cf Eqs.(\ref{6}) and (\ref{12})) are independent of the global
normalization of $\psi$.

An analogous computation can be done for the other two geometries, the plane
and the pseudosphere. When computing the averages for these two geometries, we
truncate the series in Eq.(\ref{3}) to a finite value $m = N$ and
then we let $N \rightarrow \infty$. We get 
\begin{equation} \label{7} 
\langle \psi(z) {\overline{\psi(z)}} \rangle = \left\{ \begin{array}{ll}
         (1+z {\bar z})^N \;\;\;\;\;\;\; & SU(2) \\
            \exp\left( z {\bar z} /\hbar\right) & W_1   \\
      (1-z {\bar z})^{-L} & SU(1,1) \ . \end{array} \right. 
\end{equation}

As was pointed out in Ref.\cite{kac}, the statistical properties of the
zeros of $\psi (z)$ are unaffected if instead of the distribution (\ref{pa})
a Gaussian independent distribution for the coefficients is used. This is
also true in the large $N$ limit for any statistical property of $\psi (z)$,
because the distribution (\ref{pa}) also implies in that limit a Gaussian
law for the distribution of a given coefficient. For most purposes,
Eq.(\ref{pa}) (or alternatively (\ref{1})) is equivalent to a Gaussian
independent distribution.

The random analytic model then implies that at each point of phase space the
distribution $P(\psi)$ of the function $\psi (z)$ is Gaussian, with a local
variance given precisely by Eq.(\ref{7}) (in agreement with the old
conjecture of Berry \cite{berry77})
\begin{equation} \label{15}
P\left( \psi \right) = \frac{1}{\sqrt{2 \pi \langle \left| \psi(z) \right|^2 
\rangle}} \exp\left(- \left| \psi(z)\right|^2 / 2 \langle \left| \psi(z) 
\right|^2 \rangle \right) \ .
\end{equation}
The relationship between the statistical properties of $\psi$ in different
representations was recently considered in \cite{hann2}.

\section{Ergodicity}

The simplest question concerning the statistical properties of the functions
$\psi (z)$ is related to the ergodicity. We would like to establish more
clearly the connection between the Hilbert space ergodicity of Eq.(\ref{1})
on the one hand and the phase-space ergodicity over the energy shell on the
other. From the results of the previous section this connection is now easy
to establish.

As we mention, the functions $\psi (z)$ are not stationary in the complex
$z$-plane. However, the quantity which is physically associated to the
presence of the particle in a neighborhood of a phase space cell (and
therefore more closely related to the notion of ergodicity) is not $\psi (z)$ 
but rather the (quasi) probability density
\begin{equation} \label{hus}
{\cal H}_\psi (z,{\bar z}) = \frac{|\psi (z)|^2}{\langle |\psi (z)|^2 \rangle}
\ ,
\end{equation}
usually called the Husimi function. It is simply obtained from the Wigner
transform through a Gaussian convolution. Exactly as for the average Wigner
function Eq.(\ref{2}), Eq.(\ref{hus}) implies that the average of the Husimi
function is constant in the complex $z$-plane. Because of this, the (quasi)
probability ${\cal H}_\psi d \mu$ is going to be stationary on the manifold
${\cal M}$ (and not in the complex $z$-plane where the measure is not
uniform (cf Eq.(\ref{mu})).

In particular, the previous statement applies to the distribution of the
zeros of ${\cal H}_\psi$, i.e. the density of zeros of ${\cal H}_\psi$
should be uniform over the manifold ${\cal M}$. This is going to be true for
the zeros of $\psi (z)$ as well, because both functions have the same set of
zeros \cite{note1}. In fact, the uniformity over ${\cal M}$ of the density
of zeros $\rho(z)= \delta \left[\psi(z)\right] |d\psi(z)/d z|^2$ of $\psi (z)$ 
can be directly checked for the three geometries from the  general formula
\cite{hann}
\begin{equation} \label{6}
\langle \rho(z)\rangle = \frac{1}{\pi} \frac{\partial^2}{\partial z \partial
 {\bar z}} [\ln \langle \psi(z) {\overline{\psi(z)}} \rangle ] \ .
\end{equation}
From (\ref{7}) and (\ref{6}) we find, for each geometry
\begin{equation} \label{8}
\langle \rho(z) \rangle \ d^2 z = \left\{ \begin{array}{ll}
\displaystyle \frac{N}{\pi} \frac{d^2 z}{(1+z {\bar z})^2} = \frac{N}{4 \pi}
\sin \theta \, d \theta d \varphi \;\;\;\;\; & SU(2) \\
\displaystyle   d^2 z/(\pi\hbar) & W_1 \\
\displaystyle \frac{L}{\pi} \frac{d^2 z}{(1-z {\bar z})^2} = \frac{L}{4 \pi} 
                              \sinh \tau \, d \tau d \varphi  & SU(1,1)\ .
\end{array} \right.
\end{equation}

For the sphere and the pseudosphere, the density has been transformed to the
usual coordinates spanning the corresponding manifold. Eq.(\ref{8}) shows
that the density of zeros of the random-$\psi(z)$ model is indeed uniform
over the associated manifold ${\cal M}$ (this being true for arbitrary
values of the parameters $N$, $L$ or $\hbar$). 

The uniformity of the density of zeros was also demonstrated for random
analytic functions defined on the two-dimensional torus \cite{nv}. There
exist also some related results obtained without using the random
hypothesis. For example it was proved \cite{nv} that the Schnirelman
property for the Husimi density implies a (weak) convergence of the density
of zeros towards the uniform measure. The uniformity of the distribution of
zeros was checked numerically for certain maps on the sphere \cite{bbl1,bbl2} 
and also for two-dimensional chaotic systems with $W_1$ symmetry \cite{prosen}.

In spite of its aesthetical beauty, Eq.(\ref{8}) has an unpleasant feature
for the $W_1$ and the $SU(1,1)$ groups, since a uniform density of zeros for
$\psi(z)$ over the whole (non-compact) manifold ${\cal M}$ is not realistic
from a dynamical point of view. For chaotic Hamiltonian systems, non-compact
energy shells like the plane or the pseudosphere are somewhat artificial
because energy conservation implies in general a compact manifold (or
Poincar\'e surface of section). A more appropriate model for the
eigenfunctions of a system having a $W_1$ or $SU(1,1)$ symmetry would then
be a truncated version of Eq.(\ref{3}) to a finite degree $N$. For the
plane for example, we have the $W_1$ polynomial
\begin{equation} \label{w1rp}
\psi (z) = \sum^N_{m=0} \frac{a_m}{\sqrt{m! \hbar^m}} \, z^m  \ .
\end{equation}

In practice the integer $N$ at which the truncation is made fixes the
approximate number of zeros lying in the classically allowed region of phase
space at a fixed energy (outside which the associated Husimi function has a
rapid decay). It is given by the semiclassical rule $N \approx \mbox{(area
of classically allowed region)}/ 2\pi\hbar$. The average density of roots
for this polynomial can be computed from (\ref{6}), with the result
\begin{equation} \label{9}
\langle \rho(z) \rangle = \frac{1}{\pi \hbar} \left\{ 1- g \left( \frac{z 
{\bar z}}{\hbar} \right) \left[ 1 + N - \frac{z {\bar z}} {\hbar} +
\frac{z {\bar z}}{\hbar} g \left( \frac{z {\bar z}}{\hbar} \right) \right]  
\right\} \ . 
\end{equation}
where $g(x)=(x^N /N!)/\sum\limits_{m=0}^N (x^m / m!)$. The function
(\ref{9}) has a step-like shape with a constant density of zeros
$1/(\pi\hbar$) inside a circle of radius $\sim\sqrt{N\hbar}$ and a vanishing
density outside (see Fig.2). It tends, in the large $N$-limit, to the unit
step function $\Theta(1-r)$ where $r=|z|/\sqrt{N\hbar}$. The convergence is
however slow since for large values of $r$ we find a power law dependence
$\langle \rho(r) \rangle \approx 1/(\pi \hbar N r^4) + {\cal O}(1/r^6)$.

We therefore recover for the $W_1$ random polynomials the appealing feature
that the Bargmann transform of a chaotic eigenstate has a uniform density of
zeros in a finite region of the phase space plane (over which the associated
Husimi function is stationary). The decay of the density outside the
classically allowed region is also of interest, because it describes the
evanescence of the eigenfunction $\psi$ in the classically forbidden region
\cite{tv}. But we don't expect the decay predicted by Eq.(\ref{9}) to be
very accurate due to the abrupt truncation of $\psi$ at $m=N$. 

Analogous results are obtained for the $SU(1,1)$ polynomials defined, like
the $W_1$ polynomial, by the truncation of the series in Eq.(\ref{3}) to a
finite value of $m$
\begin{equation} \label{su11rp}
\psi (z) = \sum^N_{m=0} \sqrt{C^m_{L+m-1}} \, a_m \, z^m \ .
\end{equation}
In the limit $N \gg L$ we find a constant density of zeros in the hyperbolic
plane inside a region of size $\cosh \tau \approx N/L$. 

\begin{center}
{\bf Time reversal symmetry}
\end{center}

The previous results are valid when the coefficients of the random functions
are complex. This generally corresponds to systems without time reversal
symmetry. For systems having that symmetry, the coefficients $\{ a_m \}$ can
be chosen to be real, and $\psi (z)$ satisfies now the functional equation
$$
\psi (z) = \overline{\psi ({\bar z})} \ .
$$
As discussed in detail in \cite{bbl2}, the existence of a functional
equation generically produces a concentration of a certain number $N_R$ of
roots over the symmetry line (here the real axis in the complex $z$-plane,
labeled by $x$). The distribution and correlations of roots of real
polynomials was studied in \cite{prosen1}. For $SU(2)$ polynomials, explicit
expressions for the correlations existing between real roots where obtained
in \cite{bleher}.

The number $N_R = \int \langle \rho (x) \rangle \ dx$ of real roots of a
polynomial whose coefficients are real random Gaussian independently
distributed with variance $\sigma^2_m$ may be computed from a general
formula obtained by Kac. The average density of real roots is given by
\cite{kac}
\begin{equation} \label{nu}
\langle \rho (x) \rangle = \frac{1}{\pi} \frac{\sqrt{A(x) C(x) - B^2 (x)}}{A(x)}
\ ,  
\end{equation} 
with 
\begin{equation} \label{abc} \begin{array}{ll}
A(x) = & \displaystyle \sum_{m=0}^N \sigma^2_m x^{2 m} \nonumber \\ 
B(x) = & A'(x)/2 \nonumber \\
C(x) = & [A"(x) + A'(x)/x]/4 \ , \nonumber \end{array}
\end{equation}
where the prime indicates derivative with respect to $x$.

For $SU(2)$ polynomials we have $\sigma^2_m = C_N^m$, $A(x) = (1+x^2)^N$ and
from (\ref{nu}) we obtain $\langle \rho (x) \rangle \ dx = \frac{N}{\pi}
\frac{d x}{1 + x^2}$. By a stereographic projection $x = \cot (\theta/2)$
the density simplifies to $\langle \rho (\theta) \rangle \ d\theta =
\frac{\sqrt{N}}{2\pi} d\theta$. After integration we obtain $N_R^{(SU(2))} =
\sqrt{N}$, a result which was obtained in \cite{ek}.

For real $W_1$ random functions, $\sigma^2_m = 1/(k! \hbar^k)$, $A(x) = 
\exp(x^2 /\hbar)$ and $\langle \rho (x) \rangle \ dx = d x /(\pi
\sqrt{2\hbar})$ (remember that the measure is $ d x /\sqrt{2}$). Since for
the complex $W_1$ function we have, according to Eq.(\ref{8}), $N$ roots in
a disk of radius $r=\sqrt{2 \hbar N}$, we find that the number of real roots
when the coefficients are real is $N_R^{(W_1)} = \frac{2}{\pi} \sqrt{N}$.

Finally, for $SU(1,1)$ we have $\sigma^2_m = C_{L+m-1}^m$, $A(x) =
1/(1-x^2)^N$ and $\langle \rho (x) \rangle \ dx = \frac{N}{\pi} \frac{d x}{1
- x^2}$. Going to the hyperbolic plane via $x = \tanh (\tau /2)$, the density
is simply written $\langle \rho (\tau) \rangle \ d\tau =
\frac{\sqrt{N}}{2 \pi} d\tau$. Since for the complex $SU(1,1)$ function we
have, according to Eq.(\ref{8}), a number $L$ of roots in a region of size
$\tau = \mbox{arccosh} (3)$, we find a number of real roots given by 
$N_R^{(SU(1,1))} = \frac{\mbox{arccosh} (3)}{\pi}\sqrt{N}$.

\section{Phase-space correlations -- Universality}

Besides the average density of roots, other relevant statistical properties are
correlation functions of $\psi(z)$ or of its zeros. One might think
that this quantities are geometry-dependent. However, in the large-$N$ limit,
the three different geometries treated here tend to the same flat-space result. 
Indeed, for fixed $m$ we have
\begin{equation}
\lim_{X\rightarrow\infty} C_X^m = \lim_{X\rightarrow\infty} C_{X+m-1}^m = 
\frac{X^m}{m!} \ ,
\end{equation}
where $X=N$ for the $SU(2)$ group and $X=L$ for the $SU(1,1)$. This implies
that, with the scaling $z \rightarrow \sqrt{X} z$, the $SU(2)$ and truncated
$SU(1,1)$ polynomials tend to the truncated $W_1$ polynomial in a region
around the origin of the $z$-plane (the argument can be repeated for an
arbitrary point lying inside the constant-density region). This is easy to
understand since in the large-$N$ limit the local density of zeros of the
rescaled polynomials tend to diverge (the mean spacing between zeros in
rescaled units goes like $1/\sqrt{N}$). On a scale of distances of a few
mean spacings between zeros the space will look, locally, flat. As a
consequence the {\sl local} fluctuation properties of all these functions
are, asymptotically, the same \cite{note2}. This can be rephrased as a
general statement concerning the local correlation properties of eigenstates
of fully chaotic systems (for points lying inside the energy shell): they
are {\sl universal, independent of the system under consideration or the
geometry of phase space}.

The $n$-point correlation functions among zeros have been computed
analytically by J. Hannay for random $SU(2)$ polynomials \cite{hann}. In
particular, it was found that there is a cubic repulsion between zeros. The
previous considerations imply that in the large-$N$ limit these
correlation functions should be valid locally for any geometry. The two-point
correlation function as well as a computation of the nearest neighbor
spacing distribution for the random analytic model have been compared with
numerical results obtained for several chaotic maps in different phase space
geometries in \cite{ls}. Very good agreement has been found, thus supporting
the universality of the correlations. Further support comes from a
computation of the two-point correlation function and the nearest neighbor
spacing distribution for the zeros of a two-dimensional chaotic system having
a $W_1$ symmetry \cite{prosen}.

Concerning $\psi(z)$ itself (and not its zeros), the simplest quantity of
interest is the autocorrelation function 
\begin{equation} \label{12}
C (\xi,z) = \frac{\langle \psi(z-\xi/2) {\overline{\psi(z+\xi/2)}} \rangle}
 {\sqrt{\langle |\psi(z-\xi/2)|^2 \rangle \langle |\psi(z+\xi/2)|^2 \rangle}} \ .
\end{equation}
As before, this quantity will be asymptotically independent of the geometry.
For simplicity, we restrict the computation to the $W_1$ case. Using
Eqs.(\ref{3}), (\ref{5}) and (\ref{7}) one finds
\begin{equation} \label{14}
C(\xi,z)=\exp\left[- |\xi|^2 /2 + {\rm i} \, \mbox{Im} (z {\bar \xi})\right] \ .
\end{equation}
Unlike the phase, the modulus of the autocorrelation is independent of $z$ and
is isotropic. This result should be compared with an analogous computation in
configuration space \cite{berry77}, where it was found that $C$ is given by a
Bessel function (we find that Eq.(\ref{14}) holds also for systems with time
reversal symmetry; analogous quantities for systems with and without
time-reversal symmetry have been considered more recently in \cite{prigodin}).

\section{Conclusions}

The distribution of the zeros of $W_1$ random polynomials of degree $N$ is
strongly reminiscent to that of the eigenvalues of a Gaussian ensemble of
$N\times N$ complex matrices discussed by Ginibre (see chapter 15 in
\cite{mehta}). Indeed, for large values of $N$ both distributions have a
uniform density in the complex plane inside a disk of radius $|z| \approx
\sqrt{N}$, and rapidly decaying to zero outside of it (for simplicity, we set
$\hbar = 1$ in Eq.(\ref{su11rp})). The distributions are however different.
The density has a Gaussian tail for $|z|\gg \sqrt{N}$ in the case of the
eigenvalues of random complex matrices, while it has a power law decay for
$W_1$ polynomials. Moreover, in the case of Ginibre the distribution may be
interpreted as a two-dimensional Coulomb gas, while the two-dimensional gas
associated to zeros of random polynomials includes, apart from the Coulomb
interaction, $N$-body terms \cite{ls}. A detailed comparison between these
two interacting systems was recently made in \cite{fh}.

Several extensions of the results reported here are conceivable, in
particular to higher dimensional complex manifolds. The random matrix
distribution (\ref{1}) is in fact valid independently of the number of
degrees of freedom of the underlying physical system. It is therefore
natural to expect that results analogous to those obtained here for the
distribution of the zeros in two dimensional phase spaces also hold in
higher dimensions (like for example for polynomials with several complex
variables). And indeed a generalization to $SU (m+1)$ polynomials with $m>1$
was recently done in Ref.\cite{sz}.

\vspace{0.4cm}

\noindent {\bf Acknowledgments}:
I am indebted to O. Bohigas and A. Voros for many stimulating discussions.

\newpage

\begin{center}
Captions to Figures
\end{center}

\noindent \underline{Figure 1}: (a) The sphere and, (b) the pseudosphere
(upper sheet of the hyperboloid $-x_0^2+x_1^2+x_2^2=-1$) having curvature
$+1$ and $-1$, respectively. For $SU(2)$, the definition of $z$ corresponds
to a stereographic projection which maps the unit sphere labelled by
$(\theta,\varphi)$ onto the equatorial complex plane labelled by $z$, the
north pole of the sphere being the center of projection. In the case of
$SU(1,1)$, a projection of the pseudosphere onto the unit disk in the
complex $z$-plane located at $x_0=0$ is done through the point
$(x_1,x_2,x_0)=(0,0,-1)$.\\

\vspace{2cm}

\noindent \underline{Figure 2}: The normalized density of zeros Eq.(\ref{9})
for the $W_1$ random polynomials as a function of $r=|z|/\sqrt{N \hbar}$ for
$N=100$.


\begin{thebibliography}{99}
\bibitem{note2} By local fluctuation property we mean any correlation
  function computed over phase-space distances which are of the order or
  bigger than the mean distance between zeros but small compared to the
  typical distance over which the curvature of the space is noticeable (which
  is of order one in rescaled units).
\end{thebibliography}
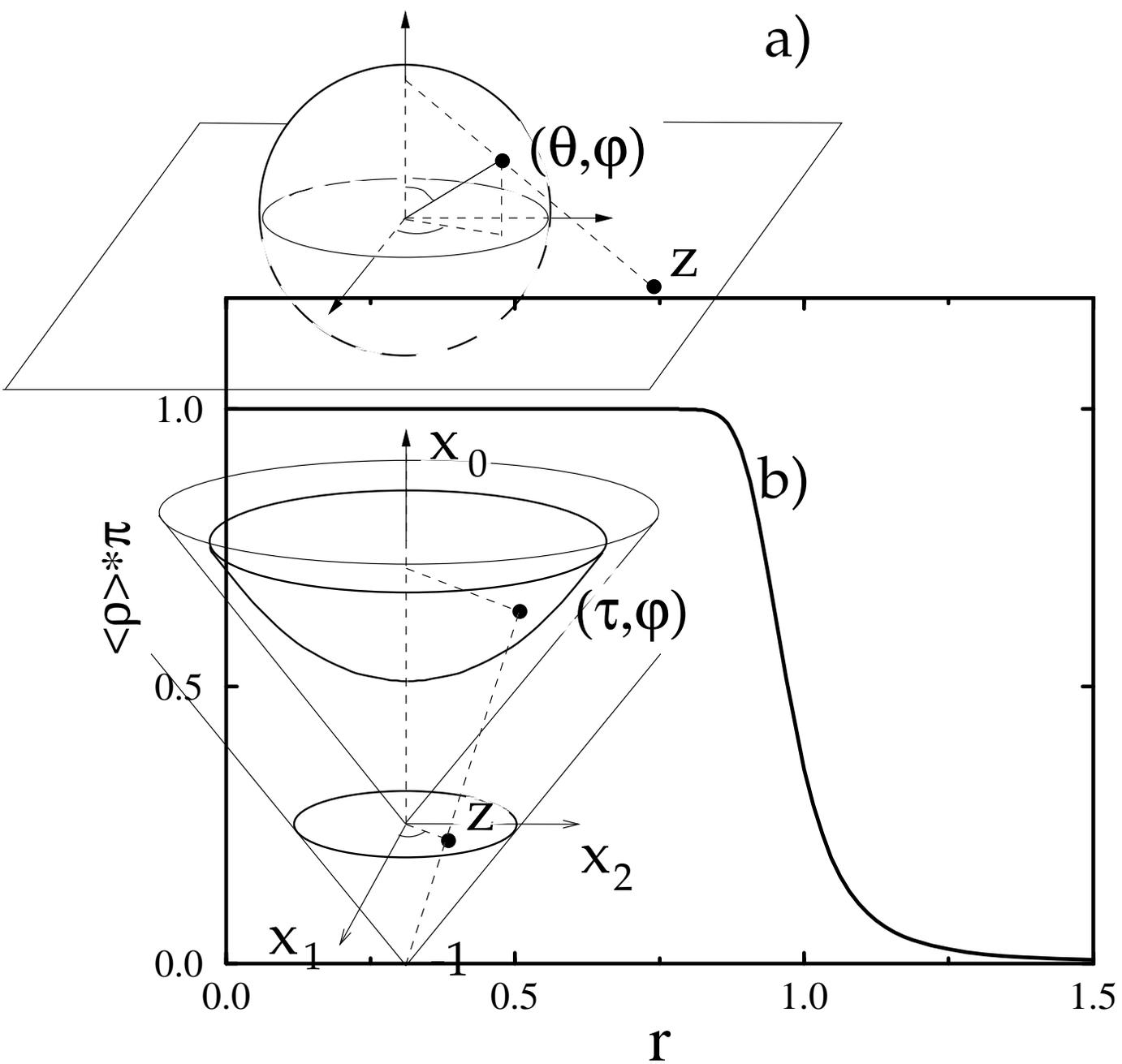
\end{document}